\documentclass[prd,amssymb,superscriptaddress,showpacs,a4paper,preprint]{revtex4-1}
\usepackage{graphicx}% Include figure files
\usepackage{dcolumn}% Align table columns on decimal point
\usepackage{bm}% bold math
\usepackage{color}
\usepackage{amsmath}
\usepackage{amssymb}
\usepackage{xcolor,cancel}

\def\vA{\vec{A}}

\def\vF{\vec{F}} \def\vPhi{\vec{\Phi}} \def\vkappa{\vec{\kappa}}
 \newcommand{\be}{\begin{equation}}
  \newcommand{\bb}{\begin{equation}}
	 \newcommand{\ee}{\end{equation}}
	 \newcommand{\ba}{\begin{eqnarray}}

		 \newcommand{\ea}{\end{eqnarray}}
		 
		   \newcommand{\bea}{\begin{eqnarray}}
			 \newcommand{\eea}{\end{eqnarray}}
\newcommand{\eqb}{\begin{eqnarray}}
\newcommand{\eqf}{\end{eqnarray}}

\begin{document}

\title{
Magnetic structures and $Z_2$ vortices in a non-Abelian  gauge model}

\author{Daniel Cabra}
\affiliation{
Instituto de F\'{\i}sica de La Plata and Departamento de F\'{\i}sica,
Universidad Nacional de La Plata, C.C. 67, 1900 La Plata, Argentina}
\author{Gustavo S. Lozano}
\affiliation{Departamento de F\'{\i}sica, FCEYN Universidad de Buenos Aires \& IFIBA CONICET,
Pabell\'on 1 Ciudad Universitaria, 1428 Buenos Aires, Argentina}
\author{Fidel A. Schaposnik$^1$\footnote{Associated to CICBA}}

\begin{abstract}
The magnetic order of the triangular lattice with antiferromagnetic interactions is described by an $SO(3)$ field and allows for the presence of $Z_2$ magnetic vortices as defects. In this work we show how these $Z_2$ vortices can be fitted into a local $SU(2)$ gauge theory. We propose simple Ans\"atzes for vortex configurations and calculate their energies using well-known results of the Abelian gauge model. We comment on how Dzyaloshinskii-Moriya interactions could be derived from a non-Abelian gauge theory and speculate on their effect on  non trivial configurations.
\end{abstract}

\maketitle

\section{Introduction}

Vortices play a central role in explaining many fundamental phenomena in Condensed Matter and High Energy Physics. The first example of vortices in theories with {\em local} gauge invariance was put forward by Abrikosov \cite{Abri}, who showed that for an intermediate range of an external magnetic fields and in a certain region of the parameter space (correponding to type II superconductors) the Ginzburg-Landau theory of superconductivity admits solutions representing a lattice of vortices.

More than 15 years later, Nielsen and Olesen discussed the role of vortices in a relativistic $U(1)$ Higgs model and pointed out its relevance in String Theory in the context of High Energy phenomena.  The first attempt to extend the study of vortices to the case of non-Abelian gauge groups can be found already in the pioneering paper of Nielsen and Olesen, who showed how to embed the Abelian solution in the $SU(2)$ non-Abelian case using two  non colinear scalar fields in the (three-dimensional) adjoint representation. It was soon realized that the correct topological characterization of these configurations implies that the topological charge is $Z_2$, unlike the abelian case where this charge is $Z$. Topologically stable non-Abelian vortex solutions were found in \cite{deVega78}-\cite{Sch}.

Many investigations followed these ideas, exploring the properties of this type of solutions for generic $SU(N)$ groups and generalizing the Yang-Mills gauge field dynamics to include Chern-Simon like terms which are important in connection with the statistics of elementary excitations \cite{Has},\cite{deVS2}.  A second wave of attention to vortices in theories with non-Abelian gauge invariance arose after  the work in Refs.~\cite{HananyTong}-\cite{SY} where the authors studied the role of
vortex-like solutions  in models that arise from the bosonic sector of ${\cal N}=2$  supersymmetric QCD with the gauge group $SU(N)\times U(1)$ and $N_f$ flavors of the fundamental matter multiplets (see \cite{SY} for a complete list of references).

Vortices also play a prominent role as excitations in magnetic systems. In a field theoretical language, these vortices correspond to non trivial configurations (defects) in theories with {\em global} gauge invariance. The best known example is that of vortices in the $XY$ model  which correspond to topologically non-trivial solutions of  $U(1)$ sigma models and play a fundamental role in the Kosterlitz-Thouless transition in two dimensional $XY$ magnetic systems.

It is also known that vortices with $Z_2$ topological charge can appear as defects in {\em antiferromagnetic} (AF) spin system in the triangular lattice since in that case the order parameter manifold is $SO(3)$   \cite{KM}. The magnetic behaviour in the AF triangular lattice can be described by {\em three} order parameters (one for each of the three sub-lattices in which the triangular lattice can be partitioned) which are themselves triplets.
In a way that resembles what happens in the $XY$ model, a Kosterlitz-Thouless transition was shown to take place, although  in the triangular $SO(3)$ case both low and high temperature phases have exponentially decaying correlations.

More recently, different studies of the triangular AF model with extra interactions, including Kitaev-like \cite{tedescos} or  Dzyaloshinskii-Moriya (DM) terms \cite{RCP}, have shown ordered phases in a magnetic field that resemble the well celebrated  $U(1)$ Abrikosov vortex lattice, but with $Z_2$ vortices instead.
%The magnetic behaviour the AF triangular lattice can be described by {\em three}
%order parameters (one for each of the three sub-lattices in which the triangular
%lattice can be partitioned) which are themselves triplets.
From a field theoretical point of view, the appearance of $Z_2$ vortices can be understood since the magnetic behaviour of the AF triangular lattice at low energies can be described by a non-linear sigma model of an order parameter field in $SO(3)$ \cite{Dombre-Read}.

The questions that naturally arise are: are these vortices allowed in a (corresponding) {\em local} gauge theory? and how are they related to the $Z_2$ vortices that were considered in the High Energy literature?

In this work we show that the  vortices of the AF magnetic system can be easily accommodated  into a local gauge theory and that, in analogy with the minimal model containing two triplets, there are two Ans\"atze that can be reduced to embeddings of the Abelian model. Using results on the energetics of vortices of the Abelian model we are able to identify the lowest energy one.

\begin{section}{$Z_2$ Vortices}
Let us start by recovering the main results of vortices in the Abelian Higgs model. The Lagrangian density describing the system is,
\be
L=-\frac{1}{4} F_{\mu \nu} F^{\mu \nu} +\frac{1}{2}(D_\mu \Phi)^{*} D^\mu \Phi-V(\Phi)
\ee
where $\Phi$ is a complex field, $F_{\mu \nu}=\partial_\mu A_\nu -\partial_\nu A_\mu$ is the electromagnetic field tensor and $D_\mu=\partial_\mu + i e A_\mu $ is the covariant derivative. Here $e$ is the charge of the scalar field (in the Ginzburg Landau theory of superconductors this is twice the electric charge). The potential can be written as
\be
V(\Phi)=c_4(\Phi^* \Phi) ^2-c_2\Phi^*\Phi
\ee
As we are working in the symmetry breaking phase, we take $c_2>0$. We work with axially symmetric vortices, so we can ignore completely the $z$-dependence of the fields. We are also interested in static solutions, so we can as well forget about the $t$-dependence. Then, we look for configurations minimizing the energy density,

\be
{\cal{E}}=\frac{1}{4} F_{ij} F_{ij} +\frac{1}{2}(D_i \Phi)^* D_i \Phi+V(\Phi)
\ee
where latin indices $i,j$ take values $x,y$.
Making the Ansatz
\be
\Phi=e^{i n \phi} f(r) \;\;\;\; A(r)=-e_\phi \frac{ a(r)}{r}
\ee
reduces the equations of motion to a system of coupled radial  second order equations. The energy (per unit length) functional becomes
\be
E_{Ab}=2\pi \int rdr  (\frac{1}{2r^2} \left(\frac{da(r)}{dr}\right)^2 + \frac{1}{2}\left(\frac{df(r)}{dr}\right)^2+
\frac{1}{r^2}\left(\vphantom{A^4 }(n+ea(r)) f(r)\right)^2+
 \frac{\lambda}{4}(f(r)^2-\eta^2)^2
 \label{energia}
\ee
where we have introduced $\lambda=4c_4$ and $\eta^2= {c_2}/(2 c_4)$.  Minimun energy configurations satisfy
\bea
\frac{d^2 a}{d r^2} -\frac{1}{r}\frac{d a}{d r}- e (n+ea) f^2 & =& 0 \nonumber \\
\frac{d^2 f}{d r^2} +\frac{1}{r}\frac{d f}{d r}-  \frac{(n+ea)^2}{r^2} f^2 + 2c_2f -4 c_4 f^2& =& 0
\ea
with boundary conditions,
\bea
& & f(0)=a(0)=0 \\
& &f(\infty)=\sqrt{\frac{c_2}{2c_4}} \\
& &a(\infty)=-\frac{n}{e}
\label{bouncon}
\ea
 When the particular relation of coupling constants $8c_4=e^2$ holds ($\lambda=e^2/2$), known as the Bogomolnyi point, this set of equations reduces to a simpler set of first order differential equations \cite{Bogo}-\cite{deVSc}. At this point,   the energy can be shown to satisfy a bound  $E=2n\pi \eta^2$. The Bogomolnyi point corresponds to the case in which the scalar mass $ m_H^2=2 \lambda \eta^2$ (inverse of the coherence length) and the vector mass $ m_v^2=e^2 \eta^2 $ (inverse of the penetration length) are equal (i.e. the limit between Type I and Type II superconductors)

The simplest  non-Abelian  extension  of the Abelian Higgs model is that in which the gauge group is $SU(2)$. The gauge fields $A_\mu$ then take values in the Lie algebra of $SU(2)$, $A_\mu = \vec A_\mu \cdot \vec\sigma /2$.  In order to have topologically stable vortices, at least $two$ non collinear scalars in the adjoint (3-dimensional) representation need to be included.  Thus we consider   scalars $\Phi_a = \vec\Phi_a\cdot \vec \sigma/2$ ($ a = 1,M$) where ($M \geq 2$),
\be
L=-\frac{1}{4} \vF_{\mu \nu} \vF^{\mu \nu} +\frac{1}{2}D_\mu \vPhi_a D^\mu \vPhi_a-V(\vPhi_a)
\label{L}
\ee
where
\be
\vF_{\mu \nu}=\partial_\mu \vA_\nu -\partial_\nu \vA_\mu + e \vA_\mu \times \vA_\nu
\ee
\be
D_\mu \vPhi_a= \partial_\mu \vPhi_a + e \vA_\mu \times \vPhi_a
\ee
The choice of the symmetry breaking potential $V(\vPhi_a)$  will be discussed below.

The $M=2$ (two-triplets) case is the best known in the literature. In this case, two possible Ans\"azte are known,
the first one, originally proposed in \cite{deVega78}, takes the form
\ba
\vPhi_1&=&f(r)(-\sin n \theta,\cos n\theta,0) \nonumber \\
\vPhi_2&=&f(r)(\cos n\theta,\sin n\theta,0) \nonumber\\
\vA_\theta&=&-(0,0,\frac{a(r)}{r})
\label{Ansatz1a}
\ea
It was later realized that another simpler Ansatz could be made \cite{deVS}-\cite{deVS2}
\ba
\vPhi_1&=&f(r)(-\sin n\theta,\cos n\theta,0) \nonumber \\
\vPhi_2&=&f(r)(0,0,1) \nonumber\\
\vA_\theta&=&-(0,0,\frac{a(r)}{r})
\label{Ansatz1b}
\ea

Although in principle one could consider arbitrary $n$, only vortices with odd $n$ are topologically non-trivial, {this corresponding to a $Z_2$ homotopy class}. Also, vortices with $n=\pm 1$ have lower energies. Moreover, it has been shown in \cite{deVS3} that vortices corresponding to Ansatz (\ref{Ansatz1b}) have lower energy, hence those associated to Ansatz (\ref{Ansatz1a}) are unstable (they will decay into the former ones).

Vortices in the Abelian Higgs model can be considered as the local gauge counterpart of the vortices of the $XY$ model, characterized by the Hamiltonian,
\be
H=-J \sum_{<ij>} \vec{S}_i \cdot \vec{S}_j
\ee
where the winding of the polar angle of the two dimensional spin $\vec{S}=S_x \vec{e}_x + S_y \vec{e}_y$ can be associated with the winding of  the complex scalar $\Phi=\Phi_1 + i \Phi_2$. Unlike the case in local gauge theories, vortices in the $XY$ model have a logarithmically divergent energy $E\thicksim \log({L}/{a})$ where $L$ represents a characteristics size of the system and $a$ is the lattice spacing.

More sophisticated vortex structures can appear in other magnetic systems. That is the case of the antiferromagnetic Heisenberg model in the triangular lattice,
\be
H=J\sum_{<ij>} \vec{S}_i \cdot \vec{S}_j
\ee
where now $\vec{S}$ is a {\em three}-dimensional vector $\vec{S}=S_x \vec{e}_x + S_y \vec{e}_y + S_z\vec{e}_z$ and ${ij}$ denote neighbors in the triangular lattice. Let us denote by $A,B,C$ the corners of a plaquette $\Delta$, then
\be
H_\Delta=J (\vec{S}_A \cdot \vec{S}_B+\vec{S}_B \cdot \vec{S}_C+ \vec{S}_C \cdot \vec{S}_A)
\ee
or
\be
H_\Delta=(\vec{S}_A + \vec{S}_B +\vec{S}_C)^2 - |\vec{S}_A|^2 - |\vec{S}_B|^2 - |\vec{S}_C|^2
\ee
Then, as the modulus of the spins are fixed, the minimum energy configuration is obtained when
\be
\vec{S}_A + \vec{S}_B +\vec{S}_C=0
\ee

\begin{figure}
\begin{center}
\includegraphics[height=2.7in]{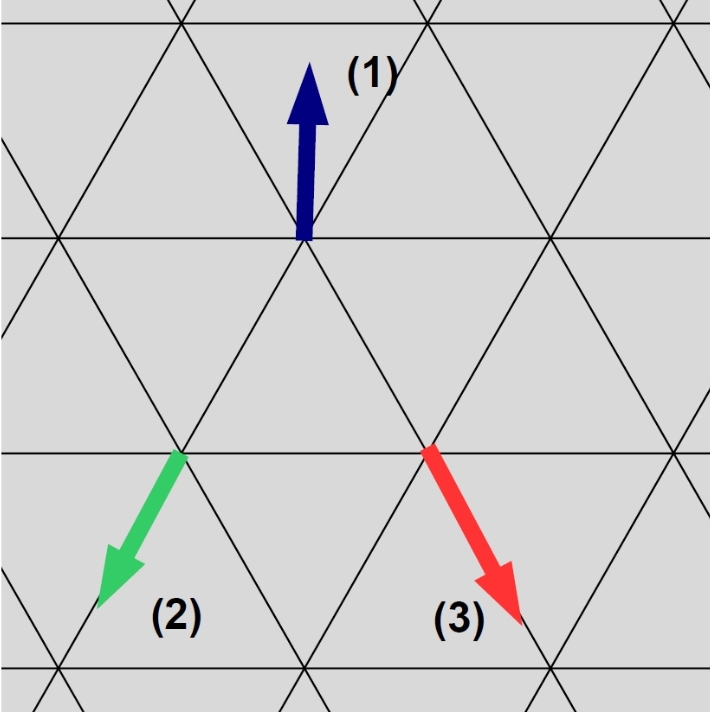}
\caption{Vacuum configuration for the AF triangular lattice. The three order parameters (denoted with different colors and numbers) are coplanar and form a $120^�$ structure in the triangular lattice}
\end{center}
\label{fig1}
\end{figure}

Thus the vacuum determines  a $120$ degrees symmetry structure of the spin configuration (see Fig~\ref{fig1}). As shown by Kawamura and S. Miyashita~\cite{KM}, vortices can appear as magnetic excitations in such systems and they are characterized by a $Z_2$ topological charge. They also discuss two possible kind of vortex configurations. In both of them, vortices are coplanar in each point, but
\begin{itemize}

\item In type I vortices, the 3 vectors are always in the same plane while they wind around the vortex center.
\item In type II vortices, one of the vectors is constant, while the other 2 wind around the vortex center.

    \end{itemize}
The energy of the vortex configurations presented in \cite{KM} where it is shown  that Type II vortices have lower energies.

Inspired by these vortices in the  antiferromgnetic triangular lattice, we consider the $SU(2)$ gauge model with three triplets field $\vec{\Phi}_a$, so $M=3$ and
$a=1,2,3$. Each field $\vec{\Phi}_a$ has three components. In the gauge theory language, these are internal indices in the Lie Algebra while in the magnetic model they refer to components in space.  In order to impose on the vacuum a $120$ degrees symmetry structure, we take a potential of the form,
\be
VC=\lambda_1 (\vPhi_1 \cdot \vPhi_1-\eta_1^2)^2 + \lambda_2 (\vPhi_2 \cdot \vPhi_2-\eta_2^2)^2+\lambda_3 (\vPhi_1 \cdot \vPhi_1-\eta_3^2)^2 + V_{mix}(\vPhi_a)
\ee
where
\be
V_{mix}(\vPhi_a)=\mu^2 (\vPhi_1 +\vPhi_2+\vPhi_3)^2
+\lambda_4 (\vPhi_1 +\vPhi_2+\vPhi_3)^4
\ee
It is clear that if we take $\lambda_i>0$, $\mu^2 >0$ and $\eta_1=\eta_2=\eta_3 \equiv \eta$,  then the vacuum corresponds to $|\Phi_i|=\eta^2$ and $\vPhi_1 +\vPhi_2+\vPhi_3=0$ (for this last condition that ensures the $120^0$ structure we do need $\mu^2 >0$). The first term in $V_{mix}$ is the analogous of the Heisenberg interaction in antiferromagnets. We have included the term with $\lambda_4$ as it is compatible with renormalization and does not change the main results of our work.
A possible vacuum configuration is illustrated  in Fig~\ref{fig1}, which is exactly the same than in the triangular lattice.

In order to find vortex configurations in the $SU(2)$ gauge model, we need to solve the field equations arising from Lagrangian \eqref{L},
\be
D^\alpha \vF_{\alpha \mu}= e D_\mu \vPhi_i \times \vPhi_i
\ee
\be
D_\mu D^\mu \vPhi_i=-\frac{\delta V}{\delta \vec\Phi_i}
\ee
%\begin{figure}
%\begin{center}
%\includegraphics[height=3.2in]{vacio1}
%\caption{A vortex configuration corresponding to the trivial vacuum.}
%\end{center}
%\end{figure}

The idea is to propose an Ansatz and determine whether  the field equations reduce to a simpler, self-consistent system of ordinary differential equations. Inspired by the (global) vortices of the antiferromagnetic triangular lattice and the vortices of the $SU(2)$, $M=2$ model described above we propose the following  Type I Ansazt,
\ba
\vPhi_1&=&f(r)(-\sin n\theta,\cos n\theta,0) \nonumber \\
\vPhi_2&=&f(r)(-\sin(n\theta+\frac{2\pi}{3}),\cos(n \theta+\frac{2\pi}{3}),0)   \nonumber\\
\vPhi_3&=& f(r)(-\sin(n \theta+\frac{4\pi}{3}),\cos(n \theta+\frac{4\pi}{3}),0) \nonumber\\
\vA_\theta&=&-(0,0,\frac{a(r)}{r})
\label{Ansatz1}
\ea
with $n \in \mathbb{Z}$.
Notice that   this Ansatz implies that
\be
\vPhi_1(r,\theta) +\vPhi_2(r,\theta) +\vPhi_3(r,\theta) =0
\label{120}
\ee

\begin{figure}
\begin{center}
\includegraphics[height=3.2in]{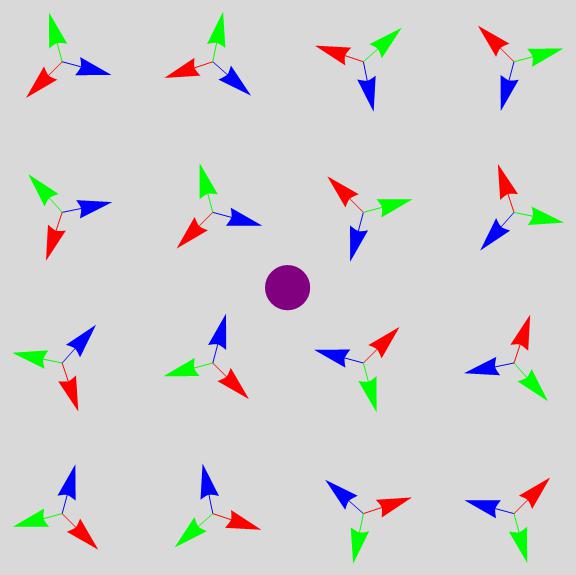}
\caption{Schematic top view of a Type I vortex. In this Ansatz the three triplets (blue, red, green) are coplanar(in the XY  plane) and wind around the core of the vortex (reprented as a disk).}
\end{center}
\label{fig2}
\end{figure}
At each point, the three triplets are then at $120$ degrees and they are always in the same plane while they wind around the origin of the vortex (the origin). The main differences with Type I magnetic vortex are of course that now we have a gauge field which we have chosen in the $3^{rd}$ direction, and that the
moduli of the triplets  are constant only at infinity, where they tend to a minimum of the potential. We have made a schematic representation of the solutions in Fig.~\ref{fig2}.

 Notice that no terms arising from $V_{mix}$ appear in the field equations since $\delta V_{mix}/\delta \Phi_i$ is a polynomial in powers of  $(\vPhi_1 +\vPhi_2+\vPhi_3)$ so that it vanishes. One then has
\be
\frac{\delta V}{\delta \vec \Phi_a} = 4 \lambda f(r)(f^2 - \eta^2) \vec \Phi_a
\ee
so that the equations of motion for the scalar fields
\be
% D_\mu D^\mu \vPhi_i=
\Box \vPhi_a-2 e \frac{a}{r} \vPhi_a-e^2 \frac{a^2}{r^2} \vPhi_a = -\frac{\delta V}{\delta \vec\Phi_a}
\ee
 reduce to the radial equation
\be
%D_\mu D^\mu \vPhi_i&=&\partial_r^2 f +\frac{1}{r}\partial_r %f-\frac{1}{r^2}f-2e\frac{a}{r^2} f-
%e^2 \frac{a^2}{r^2} f \\
 f'' +\frac{1}{r} f'-\frac{1}{r^2}(n+ea)^2f = 4\lambda f(r)(f^2 - 1)
\ee
which coincides, apart for a numerical factor in the r.h.s.,  with the  radial equation for the Abelian Higgs model scalar equation of motion.

Concerning the scalar current, once the Ansatz is inserted  it  takes  the simple form
\be
\vec J_\theta=e D_\theta \vec \Phi_a \times \vec \Phi_a=-\frac{e}{r}f^2 (n+ea) (0,0,1)
\ee
so that the gauge field radial equation of motion also reduces to the Abelian model one.

The conditions to ensure finite-energy configurations are
\[
f(0) = 0  \hspace{2cm} a(0) = 0 \]
\be
 \lim_{r\to \infty}f(r)   = \eta  \hspace {2 cm} \lim_{r \to \infty}a(r) = -\frac{n}{e}
 \label{condition}
 \ee
\begin{figure}
\begin{center}
\includegraphics[height=3.2in]{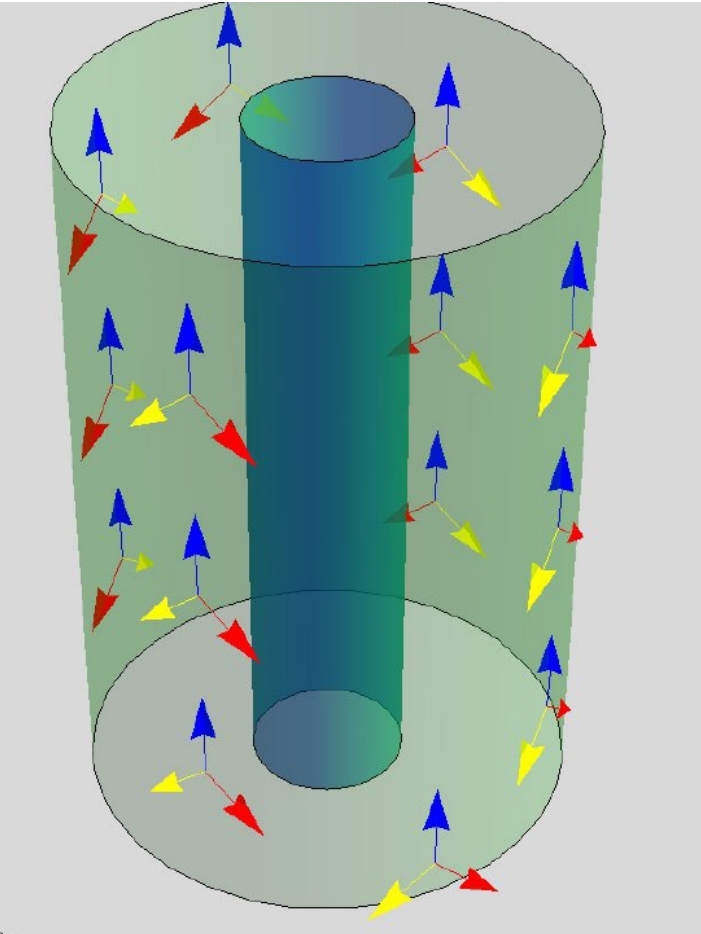}
\caption{Three dimensional view of a Type II vortex. The three triplets (blue, red, yellow) are tangent to cylinders that have the center at the core of the vortex. We have represented the region of intense chromomagnetic field with a darker color}
\end{center}
\label{fig3}
\end{figure}
Finally, the energy of static configurations satisfying the Ansatz \eqref{Ansatz1} is given by
\be
E=\int d^2 x (\frac{1}{4}\vF_{lm} \cdot \vF_{lm} + \frac{1}{2} D_l\vPhi_i \cdot  D_l\vPhi_i +V)
\ee
or
\be
E =\int d^2 x (\frac{1}{2r^2}(\partial_r a(r))^2 + \frac{3}{2}(\partial_r f(r)^2+ \frac{1}{r^2}((n+ea(r)) f(r))^2+
3 \frac{\lambda}{4}(f^2-\eta^2)^2
\label{mag}
\ee
Redefining $r=\kappa \rho$
we end up with
\be
E=2\pi \frac{1}{\kappa^2}\int  (\frac{1}{2\rho^2}(\partial_{\rho} a(\rho))^2+ \frac{3\kappa^2}{2}(\partial_\rho h(\rho)^2+ \frac{1}{\rho^2}(n+ea(\rho))^2h(\rho))^2+
 \frac{3\kappa^4\lambda}{4}(f^2-\eta^2)^2
 \label{cac}
\ee
Thus, choosing $\kappa^2=1/3$,
the energy functional which we shall denote $E^{(1)}$ becomes, apart from a factor,  identical to the Abelian one, eq.\eqref{energia}, but with $\lambda \rightarrow \lambda/3$
\be
E^{(I)}= 3 E_{Ab}(\lambda/3,e, n,\eta)
\label{E^I}
\ee

We can next take an Anstaz inspired in the Type II vortices. We then consider ,
\ba
\vPhi_1&=&(0,0,\eta) \nonumber\\
\vPhi_2&=&\frac{1}{2}(\sqrt{3}f(r)\sin(n\theta),\sqrt{3}f(r)\cos(n\theta)),-\eta)\nonumber \\
\vPhi_3&=&  \frac{1}{2}(-\sqrt{3}f(r)\sin(n\theta),-\sqrt{3}f(r)\cos(n\theta)),-\eta) \nonumber\\
\vA_\theta&=&-(0,0,\frac{a(r)}{r})
\label{AA2}
\ea

As before, Eqs \eqref{120} holds and the triplets are at $120$ degrees.
In this case the field $\vPhi_1$ does not contribute to the energy and  $D_\mu \vPhi_a$ is projected into the $(1,2)$ plane. Within this Ansatz, for each point, the triplets live in the tangent plane of a cylinder with center at the vortex core, and one of the triplets is everywhere constant (see Fig~\ref{fig3}).  As before, one can  see that inserting the Ansatz  in the equations  of motion, reduce them to a  system of coupled ordinary differential equations, which after scaling can be related to the Abelian model ones. As for the energy, plugging the Ansatz into the energy functional one obtains
\be
E=\int d^2 x (\frac{1}{2r^2}(\partial_r a(r))^2 + \frac{3}{4}(\partial_r f(r)^2+ \frac{1}{r^2}((n+ea(r)) f(r))^2+
2 \frac{\lambda}{4}(f^2-\eta^2)^2
\ee
which, after the rescaling
$
r=\kappa \rho
$
becomes
\be
E=2\pi \frac{1}{\kappa^2}\int  (\frac{1}{2\rho^2}(\partial_{\rho} a(\rho))^2+ \frac{3\kappa^2}{4}(\partial_\rho h(\rho)^2+ \frac{1}{\rho^2}(n+ea(\rho))^2h(\rho))^2+
 \frac{2\kappa^4\lambda}{4}(f^2-\eta^2)^2
 \ee
Thus, choosing,
\be
\kappa^2=2/3
\ee
one finally gets
\be
E=2\pi \frac{3}{2}\int  (\frac{1}{2\rho^2}(\partial_{\rho} a(\rho))^2+ \frac{1}{2}(\partial_\rho h(\rho)^2+ \frac{1}{\rho^2}(n+ea(\rho))^2h(\rho))^2+
 \frac{2\lambda}{9}(f^2-\eta^2)^2
 \ee
leading to
\be
E^{(II)}= \frac{3}{2}E_{Ab}(\frac{8}{9}\lambda, e,n,\eta)
\label{E^II}
\ee

%In our units,
%\be
%m_v^2=e \eta^2 \,\,\, m_H=2 \lambda \eta^2
%\ee

In order to compare the energies of these two Ans\"atzes we can borrow some results from the Abelian model. First, the energy is an increasing function of $n$. As in the $M=2$ case, there is only one class of topologically non-trivial configurations, we consider $n=\pm 1$. Second, a simple dimensional analysis shows that
\be
E_{Ab}(\lambda,e,\eta)=  \eta^2 \epsilon(\lambda/e^2)
\ee
 Now, as it is well known, for generic values of $\lambda/e^2$ there is no analytical result for  $\epsilon(\lambda/e^2)$ except at the Bogomolny point,
for which  $\epsilon(1/2)=2\pi $. For other $\lambda/e^2$ values, a numerical calculations is required. We can use the accurate result of the variational calculation presented in \cite{Hill}
 \be
\epsilon= 2.38\pi\!\!\left(\frac{\lambda}{e^2}\right)^{\!\!\alpha}
\ee
with  $\alpha=0.195...$. . Using this value for the case at hand we find
for the ratio of $E^I$ and $E^{II}$ energies as given by eqs.\,\eqref{E^I},\eqref{E^II}
\be
\frac{E^{(I)}}{E^{(II)}}=1.65....
\ee
Then, as in the global case \cite{Vilenkin} and in the $SU(2)$ gauge theory with two Higgs scalar model discussed in \cite{deVS},   the Ansatz containing one ``constant'' Higgs scalar leads to the solution having the lowest energy.

~

Note that the first term in the energy integral \eqref{cac} can be interpreted  as the
radial component of magnetic field  defined as  ,
\be
B \equiv \frac12\varepsilon_{ijk} \frac{ \vec{\Phi}_3}\eta \cdot \vec F_{jk}= \partial_ra(r)/r
\ee
 In view of the boundary conditions \eqref{bouncon} the resulting vortex magnetic magnetic flux $ {\cal F}_B$ is quantized in units of $2\pi/e$,
\be
{\cal F}_B \equiv \int d^2x B = -\frac{2\pi}e n \;,   \hspace{2 cm}  n \in \mathbb{Z}
\ee

Since the invariant group of the vacuum associated to Ans\"atzes \eqref{Ansatz1} and \eqref{AA2} is $Z_2$, the relevant homotopy group is  $\Pi_1(SU(2)/Z_2) = Z_2$. The corresponding topological charges   can be calculated via the Wilson loop
\be
Q=\frac{1}{2} {\rm Tr} \exp\left (i \oint_{C_\infty} A_\mu dx^\mu\right)
\ee
with Tr the $SU(2)$ trace and $C_\infty$ a closed curve at infinity. Both in the case of Type I and type II vortices
this gives
\be
Q=\frac{1}{2} {\rm Tr} \exp\left ({\frac{i}2 \oint_{S^1} d\theta a(r\!=\!\infty,\theta)  {\sigma_3} }\right)=\frac{1}{2} {\rm Tr}\exp\left({  i \pi n \sigma_3}\right) = (-1)^n
\ee
Hence we conclude
that there are two topologically inequivalent configurations,
the "trivial" $Q=1$ ones ($n=2k$) and those with $Q=-1$ for the "non-trivial"  ones ($n=2k+1$). Notice that the fact of being topologically non trivial does not ensure stability. Indeed we have shown that the type I Ansatz is topological non trivial but unstable towards decay into Type II Ansatz.

~

Following Kawamura and Miyashita \cite{KM} we can also define the vector chirality
\be
\vkappa=\frac{2}{3 \sqrt{3}}(\check\Phi_1 \times\check \Phi_2+\check\Phi_2 \times \check\Phi_3+\check\Phi_3 \times \check\Phi_1)
\ee
with
\be
\check \Phi_i = \frac{\Phi_i}{|\Phi_i|}
\ee
For type I vortices, this gives
\be
\vkappa=  (0,0,1)
\ee
while for type II ones one has
\be
\vkappa=  (-\cos{\theta},\sin{\theta},0)
\ee
We see that in the type I vortex the chirality vector is fixed and perpendicular to the plane where the
120 degrees structure lies while in the type II case vector $\vec \kappa$ rotates around the vortex core.

\section{Conclusions}

In this work we have analyzed vortex solutions in a non-Abelian $SU(2)$ gauge model with three matter fields in the adjoint representation (triplets) being our original motivation to determine whether the  global vortices of the triangular antiferromagnetic lattice \cite{KM} can be conveniently fitted  into a local gauge theory. We have shown that this is indeed  the case and the  vortex solutions that we have constructed share many similarities to those of the minimal theory with {\em two} triplets that have been considered for many years in the context of High Energy models.

Vortices in a local $SU(2)$ gauge theory with $M=3$ matter fields have also been recently considered in the context of QCD \cite{Oxman}. Notice however that our model and Ans\"atses are  different: in \cite{Oxman}  the triplets are perpendicular among them at each point and a different potential is chosen. Our model bears also  many similarities to those discussed in the case of three-component  superconductors, although in those systems the three order parameters are complex fields (rather than triplets) and the gauge field is Abelian \cite{Babaev}.

Coming back to the original motivation of the magnetic analogy, let us point out that the antiferromagnetic triangular lattice has been recently the focus of attention in connection to the existence of vortex and skyrmion lattices \cite{tedescos,RCP}. As shown in numerical simulations of the Heisenberg model in this lattice, the inclusion of Kitaev type and DM interactions can induce vortex and skyrmion  lattice phases in  some region of the parameter space.  In the continuum description of the Heisenberg model in the square lattice and for certain choice of the DM vectors, the DM interaction corresponds to a term in the energy of the form

\be
{\cal{E}}_M= D \epsilon_{ijk} \Phi_i \nabla_j    \Phi_k
\ee
where in the magnetic case, $\Phi_i$ correspond to the components of the order parameter living in $S^2$ \cite{Bogdanov}.

Notice that a term of this kind naturally arises in a theory with non-Abelian gauge fields. Indeed, consider the covariant derivative energy density part of the energy functional,
\be
{\cal{E}}_{cd}= | (\nabla_i\vec{\Phi} +  e \vec{A}_i \times \vec{\Phi})|^2
\ee

\be
{\cal{E}}_{cd}= | \nabla_i\vec{\Phi}|^2 +  |e \vec{A}_i \times \vec{\Phi}|^2 +2e \partial_i\vec{\Phi} \cdot (\vec{A}_i \times \vec{\Phi})
\label{mor}
\ee
The last term is
\be
2e \nabla_i\vec{\Phi} \cdot (\vec{A}_i \times \vec{\Phi})=2e  \nabla_i\Phi_l \epsilon_{lmn} {A}_{im} \Phi_n
\ee
Thus, if we choose, ${A}_{im}=\gamma \delta_{im}$ (in $A_{im}$ the first subindex denote space index and the second Lie Algebra component),
\be
2e \nabla_i\vec{\Phi} \cdot (\vec{A}_i \times \vec{\Phi})=-2e \gamma \nabla_i\Phi_l \epsilon_{nil}  \Phi_n
\ee
This is exactly the Moriya term with $D=-2 \gamma$. The second term in (\ref{mor}) is just an irrelevant quadratic factor $e^2 \gamma^2 \Phi_l \Phi_l$.

Thus, the Moriya term appears as a result of a constant $SU(2)$ background vector potential. Notice that a constant vector potential is {\em not} trivial in a
non-Abelian theory. Indeed, for our purposes it is enough to take $A_{11}=\gamma=A_{22} $, this giving a constant magnetic field $B_{33}=e \gamma^2$ which is in the third direction in the Lie Algebra and in the z-direction of space. Thus a Moriya term can be incorporated choosing a constant chromomagnetic field (it is the equivalent of the  Landau problem for a non Abelian theory). Interestingly, a similar argument is used to introduce Rashba interactions for triplets in the context of cold atoms \cite{Juze} where non-Abelian gauge fields can be engineered using laser beams.

Notice that the Moriya type term can be easily generalized to the case of a theory containing three triplets ($a=1,2,3$),
\be
{\cal{E}}_M^{(1)}= D_1 \epsilon_{ijk} \Phi_{ia} \nabla_j    \Phi_{ka}
\label{mo1}
\ee
One could also include a term of the form
\be
{\cal{E}}_M^{(2)}= D_2 \epsilon_{ijk} \epsilon_{abc}\Phi_{ia} \Phi_{jb}    \Phi_{kc}
\label{mo2}
\ee
which preserves the global $SO(3)$ invariance of the theory. Of course, one could combine these two terms in a non Abelian Chern Simons type term if one were willing to interpret $\vec{\Phi}_a$ as a vector field.

Our work could be extended in many directions. One direct generalization would be to include a non-Abelian Chern Simons term for the $\vec{A}_i$ field. Such term is interesting since it alters the statistics of excitations and binds ``chromoelectric'' charge to the vortices. Is is easy to show that the Ans\"atzes that we have presented work equally fine for this case, although the analysis of the energetics of the different Ans\"atze might require some numerical work.

Another interesting issue is to determine the role of terms like those in Eqs. (\ref{mo1})-(\ref{mo2}) in the properties of the solutions. If the magnetic analogy would carry through the local gauge theory, then one would expect that this type of terms play a fundamental role in the appearance of vortex and skyrmion lattices. We expect to report on these issues on a future work.
\end{section}

\noindent{\bf{Acknowledgments:}}
D.C.C. is financially supported by PIP-CONICET, PICT-ANPCyT and UNLP grants. F.A.S. is financially supported by PIP-CONICET,
PICT-ANPCyT, UNLP and CICBA grants. G.S.L is finacially supported by PIP-CONICET and UBA. G.S.L thanks interesting conversations with P. Tamborenea.

%\begin{figure}
%\begin{center}
%\includegraphics[height=3.2in]{vortex1}
%\caption{A vortex configuration corresponding to Ansatz 1.}
%\end{center}
%\end{figure}
%
%\begin{figure}
%\begin{center}
%\includegraphics[height=3.2in]{vortex2}
%\caption{A vortex configuration corresponding to Ansatz 2; only the components in the plane (1,2) are shown.}
%\end{center}
%\end{figure}

 \end{document}